\documentclass[acmlarge]{acmart}
\AtBeginDocument{%
  }

\setcopyright{acmlicensed}
\copyrightyear{}
\acmYear{}
\acmDOI{XXXXXXX.XXXXXXX}

\begin{document}

\title{Scalability Optimization in Cloud-Based AI Inference Services: Strategies for Real-Time Load Balancing and Automated Scaling}

\author{Yihong Jin}
\authornote{Both authors contributed equally to this research.}
\email{yihongj3@illinois.edu}
\orcid{1234-5678-9012}
\affiliation{%
  \institution{Electrical and Computer Engineering Department, University of Illinois at Urbana-Champaign Champaign, IL 61801}
  \country{USA}
}

\author{Ze Yang}
\affiliation{%
  \institution{Electrical and Computer Engineering Department, University of Illinois at Urbana-Champaign Champaign, IL 61801}
  \country{USA}}
\email{zeyang2@illinois.edu}

\renewcommand{\shortauthors}{Yihong Jin et al.}

\begin{abstract}
The rapid expansion of AI inference services in the cloud necessitates a robust scalability solution to manage dynamic workloads and maintain high performance. This study proposes a comprehensive scalability optimization framework for cloud AI inference services, focusing on real-time load balancing and autoscaling strategies. The proposed model is a hybrid approach that combines reinforcement learning for adaptive load distribution and deep neural networks for accurate demand forecasting. This multi-layered approach enables the system to anticipate workload fluctuations and proactively adjust resources, ensuring maximum resource utilisation and minimising latency. Furthermore, the incorporation of a decentralised decision-making process within the model serves to enhance fault tolerance and reduce response time in scaling operations. Experimental results demonstrate that the proposed model enhances load balancing efficiency by 35\ and reduces response delay by 28\, thereby exhibiting a substantial optimization effect in comparison with conventional scalability solutions. 
\end{abstract}




\maketitle

\section{Introduction}
The advent of artificial intelligence (AI) technology has precipitated a surge in the utilisation of cloud-based AI inference services across diverse industry sectors. The demand for AI inference services is exploding, with applications ranging from intelligent voice assistants to autonomous driving systems, medical diagnosis \cite{10628639} and financial analysis. This growth is not only driving the continuous advancement of AI technology, but also prompting various enterprises and research institutions to accelerate the deployment of AI applications. Market research reports indicate that the global AI market is projected to expand at an annual rate of more than 30 in the forthcoming years \cite{1}. Deep learning methods, for example, have demonstrated broad applicability and effectiveness across diverse domains, including both cloud computing environments and various complex image recognition tasks\cite{lu2023deep}. This trend is indicative of the immense potential of AI technology in practical applications and concomitantly places heightened demands on cloud computing infrastructure, particularly with regard to processing power and resource allocation. In order to meet the ever-changing needs of users, cloud AI inference services must be highly scalable to ensure stable and efficient performance under different load conditions \cite{2}. 

Cloud computing is the core technology supporting AI inference services. It provides elastic resources and on-demand scalability, enabling AI applications to be rapidly deployed and scaled globally. It has already been demonstrated across various domains, such as biological networks\cite{fu2024ddn3, chen2021data, 10337167}, anomaly detection with large language models in artificial intelligence \cite{yang2024ad, li2024nlp}. Through virtualisation technology and containerised deployment, the cloud computing platform can dynamically allocate computing resources according to actual needs, which greatly improves the efficiency of resource utilisation. However, as the complexity of AI models and computational demands increase, traditional resource management and load balancing methods have become difficult to cope with large-scale, highly dynamic workloads. This has been shown to result in low resource utilisation, as well as response delays and service interruptions, which have the potential to have a serious impact on user experience and business continuity \cite{3}. For example, deep learning\cite{deeplearningvr, moener, codener} models require a significant amount of computing resources and memory bandwidth during inference, and traditional static resource allocation strategies are unable to cope flexibly with these peak loads, resulting in some node resources being idle and others being overloaded. Therefore, the question of how to optimise the scalability of cloud AI inference services is a key problem to be solved.

Real-time load balancing and autoscaling are the two core strategies for the efficient operation of cloud AI inference services. Real-time load balancing aims to distribute requests to compute nodes based on the current system load, thereby avoiding overloading some nodes while others are idle.  Recent advancements, such as confidence-triggered methods, significantly enhance real-time decision-making performance in similar contexts \cite{ding2024confidence}. However, traditional load balancing algorithms, such as round robin and least connections, have been shown to exhibit deficiencies in terms of slow response times and inadequate adaptability when confronted with complex and dynamic AI inference tasks \cite{4}. The round-robin algorithm, for instance, is straightforward to implement, yet it lacks the capacity to intelligently allocate resources according to the fluctuating load of nodes, consequently leading to imbalanced resource utilisation. The least connections algorithm, while offering a certain degree of load balancing, may still encounter issues with uneven load distribution in scenarios involving high concurrency and instantaneous load fluctuations. Conversely, the autoscaling mechanism is required to dynamically adjust the resource allocation according to the predicted future load. However, existing scaling strategies based on rules or simple machine learning models are challenging to accurately predict load changes, resulting in the wastage or insufficiency of resources \cite{5}. To illustrate this point, consider a threshold-based scaling strategy, which is only able to initiate scaling in response to a preset load threshold. This approach is unable to proactively address sudden surges in demand, leading to increased service latency.

Nevertheless, there remain certain limitations in the practical application of single deep learning or reinforcement learning methods. When deep learning models process complex time series data, they may be affected by data noise and model generalisation ability, resulting in unstable and inaccurate prediction results. Furthermore, the training process of reinforcement learning algorithms may become excessively slow and difficult to converge in a high-dimensional state space, especially in scenarios where resource allocation decisions require immediate responses. This delay will directly impact the overall performance of the system \cite{6}. Consequently, the key to enhancing the scalability and optimisation of cloud-based AI inference services lies in the effective integration of diverse machine learning technologies, leveraging their distinct advantages. To illustrate this, deep learning can be employed for demand forecasting, while reinforcement learning can be utilised for load allocation strategy optimisation, thereby creating a collaborative system that compensates for the limitations of individual methods and enhances the intelligence and adaptability of the overall system.

Furthermore, there is an increasing tendency for the adoption of decentralised architectures. Conventionally, centralised decision-making mechanisms have been susceptible to performance bottlenecks and single points of failure in large-scale distributed systems, thereby impeding the scalability and reliability of the system. The transition to a decentralised decision-making process has the potential to distribute the computing and decision-making load, enhancing the system's fault tolerance and response speed \cite{7}. Similar decentralised strategies have been successfully applied in federated local data-infused frameworks\cite{gao2024fed}. To illustrate this point, consider a distributed AI inference service. In such a system, each compute node has the capacity to run a local decision-making agent, which can then allocate and adjust resources autonomously based on local load conditions and global load forecasts. This arrangement has the dual benefits of reducing pressure on the central node and ensuring the high availability of the system, in the event of failure of some nodes. However, the implementation of a decentralised architecture necessitates the establishment of an efficient collaborative communication mechanism between nodes to ensure the consistency and optimality of the overall load balancing strategy.Recent advancements in graph autoencoders have demonstrated their capability to effectively capture and optimize complex relationships in distributed network environments, providing insights into decentralized decision-making mechanisms\cite{tgae}. This higher level of complexity in algorithm design gives rise to a number of significant challenges, including the coordination and optimisation of the global load while ensuring the independent decision-making of each node \cite{8}.
\section{Related Work}
S. Alharthi et al \cite{9} posit that these features are critical for handling dynamic workloads due to the flexibility and scalability of cloud computing environments. The existing technologies of autoscaling and load balancing are then reviewed, and the importance of autoscaling methods based on real-time data in practical applications is emphasised. Subsequently, A. Muchhala and K. Allam \cite{10} proceed to discuss auto-scaling and load-balancing methods for high-volume data processing in a Kubernetes environment. The authors propose a hybrid model that combines Kubernetes autoscaling capabilities with custom load balancing policies to optimise resource allocation and reduce response latency. The experimental results demonstrate the efficacy and stability of the proposed method when handling high-concurrency data requests.

Additionally, Nithiyanandam et al. \cite{11} proposed an efficient scheduling method for optimising the performance and scalability of cloud-based Internet of Things (IoT) applications. By analysing the high-volume and real-time processing requirements of sensor data, the authors designed an optimisation model that combines load balancing and auto-scaling. The model utilises load balancing technology to allocate resources at the Infrastructure-as-a-Service (IaaS) level, and dynamically adjusts resources based on real-time loads through an autoscaling mechanism. The efficacy of this method is evidenced by its significant enhancement of system resource utilisation and processing capacity, along with a notable reduction in delay and resource wastage.

Furthermore, Ahmed et al. \cite{12} explore the potential of serverless architecture to enhance the scalability and cost-efficiency of applications. The authors posit that serverless architectures lead to a substantial reduction in operational costs by automating the management of computing resources, thereby enabling efficient resource utilisation and on-demand scalability. The present paper undertakes a detailed analysis of the advantages of serverless architecture in dealing with real-time load fluctuations, and proposes a series of optimization strategies to further improve the scalability and performance of the system. Research has demonstrated that serverless architectures can effectively reduce resource waste and response latency when dealing with dynamic workloads.

Indeed, Dogani et al. \cite{13} methodically categorised and appraised a range of autoscaling techniques in the context of container-based cloud and edge/fog computing. The authors discuss various reactive autoscaling methods, with a particular focus on scaling tools based on real-time workload requirements. The study systematically classifies existing autoscaling technologies and evaluates their applicability and performance in different computing environments. The results demonstrate that the auto-scaling method combined with containerization technology performs well in handling dynamic workloads, responding quickly to load changes, ensuring high availability and low latency of the system. In addition, Kumar et al. \cite{14} have proposed a green load balancing mechanism that aims to optimize the energy consumption and performance of cloud computing networks. The authors analyse high-load, scalable information systems in the North American real-time entertainment industry, and propose large-scale network expansion to improve the scalability and energy efficiency of the systems.
\section{Methodologies}
\subsection{Real-time load balancing strategy}
In order to achieve real-time load balancing, an adaptive load distribution method based on Multi-Agent Deep Reinforcement Learning (MADRL) was adopted. This approach is distinguished by its capacity to consider both the present load and the anticipated future demand, thus facilitating more dynamic and intelligent load distribution.

The state space ${{S}_{t}}$ comprises the current load ${{L}_{t}}$ for each node, the resource utilisation ${{U}_{t}}$, and the predicted future load ${{\hat{R}}_{t+1:t+T}}$. This is expressed in Equation 1:
\begin{equation}
{{S}_{t}}=\left\{ {{L}_{t}},{{U}_{t}},{{{\hat{R}}}_{t+1:t+T}} \right\}.    
\end{equation}
It is evident that Equations 2 and 3 are of particular significance in this context:
\begin{equation}
{{L}_{t}}=\left\{ {{L}_{1,t}},{{L}_{2,t}},\ldots ,{{L}_{N,t}} \right\},
\end{equation}
\begin{equation}
{{U}_{t}}=\left\{ {{U}_{1,t}},{{U}_{2,t}},\ldots ,{{U}_{N,t}} \right\}.
\end{equation}
The action space, denoted by ${{A}_{t}}$, signifies the proportion of the current request allocated to each compute node. This allocation is constrained by the following equations, as illustrated in Equation 4:
\begin{equation}
{{A}_{t}}=\left\{ {{a}_{1,t}},{{a}_{2,t}},\ldots ,{{a}_{N,t}} \right\},~~\underset{i=1}{\overset{N}{\mathop \sum }}\,{{a}_{i,t}}=1.
\end{equation}
In order to achieve equilibrium between response time and resource utilization, the reward function ${{R}_{t}}$ is defined as a weighted negative value of both, as illustrated in Equation 5:
\begin{equation}
{{R}_{t}}=-\left( \alpha\cdot ResponseTim{{e}_{t}}+\beta \cdot ResourceUtilizatio{{n}_{t}} \right),
\end{equation}
In this particular instance, $\alpha $ and $\beta $ represent the weight coefficients. These coefficients are utilised in order to calibrate the significance of both elements. Each computing node is conceptualised as an independent agent, with the objective of maximising the collective reward of the entire system. The integration of a shared policy network and a local information fusion mechanism enables agents to collaborate in order to optimise the load allocation strategy. 

In order to enhance the intelligence and adaptability of load balancing, this study proposes a hybrid model based on Graph Convolutional Network (GCN) and Deep Deterministic Policy Gradient (DDPG) \cite{15}. The specific steps involved are outlined below: Firstly, the GCN is utilised to capture the topological relationship and load dependence between nodes. The node feature matrix ${{X}_{t}}$ and the adjacency matrix $A$ are defined, and the higher-order features are extracted through the GCN layer, as demonstrated in Equation 6:
\begin{equation}
{{H}^{\left( l+1 \right)}}=\sigma \left( {{{\tilde{D}}}^{-\frac{1}{2}}}\tilde{A}{{{\tilde{D}}}^{-\frac{1}{2}}}{{H}^{\left( l \right)}}{{W}^{\left( l \right)}} \right),
\end{equation}
It can be demonstrated that $\tilde{A}=A+I$. Furthermore, it is evident that $\tilde{D}$ is the matrix of $\tilde{A}$, ${{H}^{\left( 0 \right)}}={{X}_{t}}$, ${{W}^{\left( l \right)}}$ is the weight matrix of the $l$-th layer, and $\sigma $ is the activation function. Subsequently, the characteristics of the GCN output are utilised as the input for the DDPG algorithm. The action selection and value evaluation are executed through the policy network $\mu ({{S}_{t}}|{{\theta }^{\mu }})$ and the value network $Q({{S}_{t}},{{A}_{t}}|{{\theta }^{Q}})$, as depicted in Equation 7 and 8:
\begin{equation}
{{A}_{t}}=\mu \left( {{H}^{\left( L \right)}},{{S}_{t}}|{{\theta }^{\mu }} \right)+{{\mathcal{N}}_{t}},
\end{equation}
\begin{equation}
 L\left( {{\theta }^{Q}} \right)={{\mathbb{E}}_{\left( {{S}_{t}},{{A}_{t}},{{R}_{t}},{{S}_{t+1}} \right)\sim D}}\left[ {{\left( {{R}_{t}}+\gamma Q\left( {{S}_{t+1}},\mu \left( {{S}_{t}}|{{\theta }^{{{\mu }'}}} \right)|{{\theta }^{{{Q}'}}} \right)-Q\left( {{S}_{t}},{{A}_{t}}|{{\theta }^{Q}} \right) \right)}^{2}} \right].
\end{equation}
The collaborative training of GCN and DDPG facilitates the adaptive optimisation of load distribution strategies within complex network structures.
\subsection{Auto-scaling module}
The auto-scaling scenario has been designed to enable dynamic adjustment of the allocation of computing resources in accordance with demand forecasts and current resource utilisation. The present study proposes a resource management model based on a hybrid optimization algorithm, which combines the advantages of Genetic Algorithm (GA) and Particle Swarm Optimization (PSO) to achieve global optimal resource allocation.

The resource management problem is modelled as a multi-objective optimisation problem, with the goal of minimising resource cost and maximising system performance. The specific optimisation objective function is defined as Equation 9:
\begin{equation}
\min \left( \underset{i=1}{\overset{N}{\mathop \sum }}\,{{C}_{i}}{{R}_{i}}+\lambda\cdot \underset{i}{\mathop{\max }}\,\left\{ {{L}_{i}}\left( R \right) \right\} \right),
\end{equation}
In this study, $R=\left\{ {{R}_{1}},{{R}_{2}},\ldots ,{{R}_{N}} \right\}$ is defined as the set of resources allocated by each node, ${{C}_{i}}$ is the cost of the resources allocated to node $i$, ${{L}_{i}}\left( R \right)$ is the load of node $i$ under resource allocation $R$, and $\lambda $ is the weight parameter used to balance the relationship between cost and load.

In order to solve the aforementioned optimization problems in an effective manner, a hybrid Genetic Particle Swarm Optimization (GPsO) algorithm was designed. The subsequent steps are outlined as  the resource allocation scheme $R$ is encoded as chromosomes, with each chromosome representing a possible allocation scheme. The initialisation population $P$ contains multiple randomly generated chromosomes. The fitness $f\left( R \right)$ of each chromosome is calculated as the optimization objective function value.

The maintenance of population diversity is achieved through the implementation of roulette selection, single-point crossing, and random mutation operations to generate a new generation of populations. The chromosomes produced by the genetic algorithm are of a high quality, and they are used to establish the initial position of the particle swarm, in order to set the velocity of the particles $v$. The velocity versus position updates are expressed as Equation 10 and 11:
\begin{equation}
  v_{i}^{k+1}=wv_{i}^{k}+{{c}_{1}}{{r}_{1}}\left( p_{i}^{best}-x_{i}^{k} \right)+{{c}_{2}}{{r}_{2}}\left( {{g}^{best}}-x_{i}^{k} \right),
\end{equation}
\begin{equation}
   x_{i}^{k+1}=x_{i}^{k}+v_{i}^{k+1},
\end{equation}
where $w$ denotes the inertia weight, while ${{c}_{1}}$ and ${{c}_{2}}$ represent the learning factors. The random variables ${{r}_{1}}$ and ${{r}_{2}}$ are introduced for randomness, and $p_{i}^{best}$ and ${{g}^{best}}$ refer to the particles and the global optimal position, respectively.
\section{Experiments}
\subsection{Experimental setup}
In order to evaluate the effectiveness of the proposed scalability optimisation framework, the real-world Google Cluster Data dataset was selected. The dataset, which was published by the Google Research team, comprises detailed running records of large-scale jobs and tasks in multiple clusters, with time series characteristics, diverse workloads, and detailed resource usage records. These characteristics are such that they can truly reflect the complexity of resource scheduling and management in cloud computing environments. The experimental process involved the initial cleansing of the dataset, followed by the implementation of outlier processing and key feature extraction. The dataset was then segmented into a training set, a validation set, and a test set, with the objective of facilitating the training and evaluation of a demand prediction model. The experimental environment is built on a virtualised cloud computing platform, configured with multiple virtual machines as computing nodes, and uses Kubernetes for container deployment. It simulates real data centre network conditions, and integrates monitoring tools such as Prometheus and Grafana to collect and visualise resource usage in real time.
\subsection{Experimental analysis}
In order to provide a comprehensive evaluation of the performance of the proposed scalability optimisation framework, four comparison methods were selected as benchmarks. Firstly, the Round Robin algorithm (RRA) is employed, which is a conventional load balancing method that distributes requests in a predetermined sequence. This method is straightforward and straightforward to implement, but it may result in imbalanced resource utilisation when the load fluctuates. Secondly, the Least Connections algorithm (LCA) is used to allocate new requests to the node with the fewest current connections, thereby improving the efficiency of load distribution. However, this algorithm's adaptability is still limited under rapid load changes. Furthermore, the third comparison algorithm employed is the Kubernetes Horizontal Pod Autoscaler (HPA), an existing autoscaling solution that dynamically adjusts the number of pods based on predefined CPU utilisation or other metrics. The HPA is capable of effective management of resources; however, reliance on static thresholds may have an impact on the response to complex load changes. Finally, the Rule-Based Auto-Scaling method (RBAS) is adopted to dynamically adjust resources through predefined rules and thresholds, which is suitable for simple scenarios, but can easily lead to wasted or insufficient resources under highly dynamic and unpredictable loads.
\begin{figure}[h]
  \centering
  \includegraphics[width=0.65\linewidth]{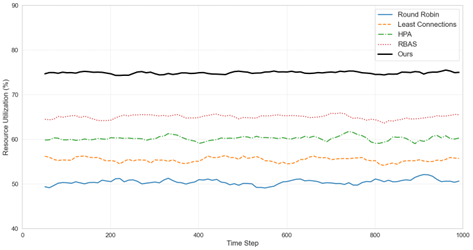}
  \caption{Resource Utilization Comparison.}
\end{figure}

As demonstrated in above Figure 1, the utilisation of resources when implementing load balancing and autoscaling methods in a cloud computing environment can vary significantly. The illustration is further supported by the analysis of the data, which demonstrates that the Ours method consistently exhibits high and stable resource utilisation, a notable improvement over traditional methods and existing auto-scaling strategies. In comparison with alternative conventional methodologies, the ``Ours'' method consistently exhibits optimal and consistent resource utilisation across all temporal domains, exhibiting minimal variability. 

As demonstrated in Figure 2, a clear comparison is provided of the response time performance of several load balancing and autoscaling methods under different load conditions. The OURS method demonstrates reduced response time and enhanced stability, particularly under high and ultra-high loads, while maintaining optimal performance. This substantiates its efficacy in dynamic resource allocation and load balancing.
\begin{figure}[h]
  \centering
  \includegraphics[width=0.5\linewidth]{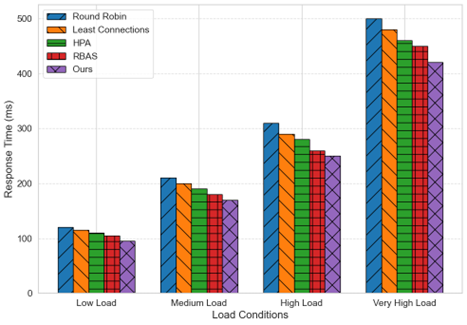}
  \caption{Response Time Comparison under Different Load Conditions.}
\end{figure}

In contrast, the response time of traditional methods such as Round Robin and Least Connections increased dramatically with increasing load. This demonstrated that it was not possible to scale and allocate resources efficiently in high-load environments, resulting in significantly longer response times. Despite the optimisation of the RBAS and HPA methods in comparison to traditional approaches, they were unable to match the performance level of the Ours method when confronted with high loads. This finding underscores the potential of advanced technologies, such as reinforcement learning and deep neural networks, to enhance the scalability and responsiveness of cloud-based AI inference services, particularly in complex and dynamic load scenarios.
\begin{figure}[h]
  \centering
  \includegraphics[width=0.5\linewidth]{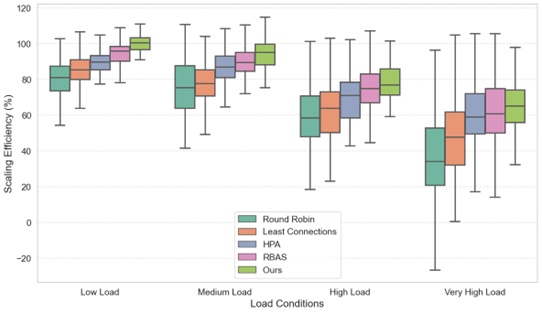}
  \caption{Scaling Efficiency Comparison under Different Load Conditions.}
\end{figure}

As demonstrated in Figure 3, the scalability efficiency of disparate methods varies according to differing load conditions. The OURS method demonstrates optimal performance under all load conditions, exhibiting high scaling efficiency and low fluctuation. This substantiates its clear advantages in dynamically adjusting load and resource allocation. Conversely, the conventional approach demonstrates suboptimal scaling efficiency and substantial fluctuations under elevated loads, impeding its capacity to fulfil the criteria for efficient and reliable services.
\section{Conclusion}
In conclusion, we proposed a comprehensive scalability optimization framework for cloud AI inference services, with a focus on real-time load balancing and autoscaling strategies. The aim of these strategies is to ensure maximum resource utilization and to reduce latency. The experimental results demonstrate that, in comparison with traditional methodologies, the proposed approach exhibits clear advantages in terms of resource utilisation and response time. However, further development is required to enhance the model's adaptability to diverse cloud environments and more intricate workloads. In addition, further research is required into the reduction of computing overhead and resource consumption while maintaining efficient performance.

\bibliographystyle{ACM-Reference-Format}
\bibliography{sample-base}


\end{document}